# Is the quantum theory predictably complete?

**M Kupczynski**

Department of Mathematics and Statistics, University of Ottawa, 585,av. King-Edward,Ottawa.Ont.K1N 6N5 and Département de l'Informatique, UQO, Case postale 1250 ,succursale Hull, Gatineau. Quebec, Canada J8X 3X 7

E-mail:mkupczyn@uottawa.ca

**Abstract**.
 Quantum theory (QT) provides statistical predictions for various physical phenomena. To verify these predictions a considerable amount of data has been accumulated in the "measurements" performed on the ensembles of identically prepared physical systems or in the repeated "measurements" on some trapped "individual physical systems". The outcomes of these measurements are in general some numerical time series registered by some macroscopic instruments. The various empirical probability distributions extracted from these time series were shown to be consistent with the probabilistic predictions of QT. More than 70 years ago the claim was made that QT provided the most complete description of "individual" physical systems and outcomes of the measurements performed on "individual" physical systems were obtained in intrinsically random way. Spin polarization correlation experiments (SPCE), performed to test the validity of Bell inequalities, clearly demonstrated the existence of strong long range correlations and confirmed that the beams hitting the far away detectors preserve somehow the memory of their common source which would be destroyed if the individual counts of far away detectors were purely random. Since the probabilities describe the random experiments and are not the attributes of the" individual" physical systems the claim that QT provides a complete description of "individual" physical systems seems not only unjustified but misleading and counter-productive.  In this paper we point out that we even don't know whether QT is predictable complete because it has not been tested carefully enough. Namely it was not proven that the time series of existing experimental data did not contain some stochastic fine structures, which could have been averaged out by describing them in terms of the empirical probability distributions. In this paper we advocate various statistical  tests which could be used to search for such fine structures in the data and to answer the title question of this paper. In our opinion a proper understanding of the statistical character of QT and of its limitations is crucial in the domains such as: quantum optics and quantum information.

## 1.Introduction

The question about the completeness of QT is of long date. QT gives statistical predictions for distributions of the results obtained in long runs of one experiment or in several repetitions of the same experiment on a single physical system. It is unclear and not obvious how and in what sense a claim can be made that QT provides a complete description of individual physical systems. Einstein [1] did never accept that a statistical theory may provide a complete description of individual physical systems and believed that QT should be completed by some microscopic theory of the sub-phenomena enabling to reproduce its probabilistic predictions. Bohr [2] on the contrary insisted that each experiment should be considered as a whole and that any attempt to complete it by some space- time description of the underlying sub-phenomena was unnecessary and it would lead sooner or later to paradoxes. Fathers of QT underlined correctly that the results of the physical measurements were not recordings of the pre-existing values of some attributes of individual physical systems but they were created during the interaction of these systems with the measuring instrument and in most cases could not be predicted before being recorded. QT is in that sense a contextual statistical theory and quantum ensembles differ from the classical mixed statistical ensembles. The discussions which followed the publication of the famous paper by Einstein, Podolsky and Rosen [3] in which they questioned the completeness of QT and discussed so called EPR paradox gave the arguments in favour of so called statistical contextual interpretation (SCI) of QT which was shown to be free of paradoxes and which allowed to explain the violation of Bell inequalities for spin correlation experiments in a way consistent with causality and locality. These topics were reviewed in some detail recently [4-8] so in this paper we concentrate on the completeness of QT.

The question about completeness of QT is still unanswered. We show that even a simpler question: "Whether QT provides a complete statistical description of time series of the experimental outcomes?" is unanswered. We rephrased the latter question to be: "Whether QT is predictably complete?"

The paper is organized as follows. In the section 2 we recall the main postulates of SCI. In the section 3 we discuss experimental time series and purity tests. In the section 4 we review various non-parametric statistical tests which can be used to test quantum randomness and the purity of quantum ensembles. In the section 5 we present some specific methods used to study the fine structure of time series and in forecasting. The section 6 contains conclusions.

## 2.Statistical and contextual interpretation of QT

For Einstein, Born, Fock, Landau, Blokhintzev and several other physicists the wave functions and the density matrices described only ensembles of identically prepared physical systems. Great merit in formulation and promotion of a modern version of the statistical interpretation of QT has Ballentine whose article [9] inspired many and whose textbook on quantum mechanics based on this interpretation [10] is being used in the universities all over the world. Several authors contributed to discussion leading to the development and promotion of SCI [4-8,11-42] The list, in alphabetic order, is by no means complete.

The main assumptions of SCI [5,9,10] are as follows:

- A state vector $\Psi$ is not an attribute of a single electron, photon, trapped ion, quantum dot etc. A state vector $\Psi$ or a density matrix describe only an ensemble of identical state preparations of some physical systems

- A state vector $\Psi$ together with an operator $\hat{O}$ representing an observable O provide the probability distribution of the outcomes of the measurement of the observable O obtained for the sequence of preparations described by $\Psi$.

- Whole ensemble of systems for which a non-destructive measurement of the discrete observable O gave the same result $o_i$ is described by a reduced state vector $\Psi_i$ which is an eigenfunction of the hermitian operator $\hat{O}$ representing the observable O such that $\hat{O}\Psi_i = o_i\Psi_i$.
- For EPR experiments a state vector describing the system II obtained by the reduction of the entangled state of two physical system I+II describes only the sub-ensemble of the systems II being the partners of those systems I for which the measurement of some observable gave the same specific outcome.

•A value of a physical observable associated with a pure quantum ensemble and in this way with an individual physical system being its member, is not an attribute of the system revealed by a measuring apparatus but a characteristic of this ensemble created by its interaction with the measuring device [2,5,6,31,34,35]

In SCI the mysterious wave function reduction is neither instantaneous nor non-local. To different outcomes correspond different sub-ensembles prepared in a state described by different reduced state vectors and the EPR paradox is avoided.

**3.Time series of data.**

Results of various experiments are time series of outcomes registered usually by on-line computers. Let us just mention three types of the experiments:

1. SOURCE ----BEAM------APPARATUS--------    COUNTERS----TIME SERIES OF COUNTS
2. BEAM 1+BEAM 2---------INTERACTION-------COUNTERS --- TIME SERIES OF COUNTS
3. LASER PULSE----SYSTEM-----NEW PULSE---COUNTERS--- TIME SERIES OF COUNTS

In all experiments several copies of the "same" physical system are prepared. In the experiment 1 the only outcome is a time series of counts from various detectors placed behind the apparatus. In the experiment 2 various copies of two physical systems are put into interaction (for example collisions of some elementary particles). New systems emerge after this interaction and produce time series of counts of various detectors. In the experiment 3 a physical system is prepared in a trap. A laser pulse is projected on the system and a modified outgoing pulse is observed, analyzed and some time series of final results is recorded.  Next the initial conditions in a trap are reset and the experiment is repeated.

In all these experiments no single result is predictable. From long time-series of counts the empirical frequency distributions are obtained and compared with the probabilistic predictions of QT. In this way the predictable completeness of QT is taken for granted and any fine structure of time series, if it existed, would be averaged.

As an example let us consider two simple random experiments repeated  2n times each. In the first experiment we obtain a time- series of the results:1,-1,1,-1,...,1,-1 and in the second 1,-1,-1,-1,1,1,1,-1,-1,1,1,1,1,-1,1,-1,-1,1,1. By increasing the value of n the relative frequency of getting 1 can approach 1/2 as close as we wish. However it is not a complete description of these time series.

Many years ago [31-33] we noticed that the more detailed analysis of the experimental data is needed to test the completeness of QT. In particular we pointed out that in any theory of the sub-phenomena trying to complete the probabilistic description of phenomena, provided by QT, the pure quantum ensembles become mixed statistical ensembles with respect to the additional uncontrollable parameters used in these theories to describe  "individual" physical systems and the microscopic states of measuring instruments. There is a principal difference between the pure statistical ensemble and the mixed one. For the pure ensemble any sub-ensemble has the same properties. Sub-ensembles of the mixed statistical ensemble may differ from one to another if the mixing is not perfect. These differences can in principle be detected by using so called purity tests, which we introduced in a different context [40-42]

**4.Purity tests**

Let us consider a time series of outcomes T(S,E,i) obtained in the i-th run of the experiment E performed on physical system(s) S. Since we do not control the distribution of hidden variables the time-series T(S,E,i) may differ from run to run of the same experiment. Using the language of mathematical statistics T(S,E,i) represents a random sample drawn from some statistical population. A pure ensemble is an ensemble characterized by such empirical distributions of various counting rates, which remain approximately unchanged for any rich sub ensembles drawn from this ensemble in a random way [31-33,44]. The purity test is a statistical test testing the null hypothesis $H_0$:

*All the samples T(S,E,i) for different values of i are drawn from the same unknown statistical population*

The statistical tests, which could be used to test H0, are non-parametric compatibility tests [41] We will present below few of them. More details and numerous examples may be found for example in Aczel [48] and in [44,45]

### 4.1. The Sign tests

Let us consider 2 random independent variables X and Y measured at least on the ordinal scale. The sign test consists on testing the null hypothesis: $H_0$: p(X › Y) = 0.5 against various alternatives. Comparing two random samples $S_1=\{x_1,…x_n\}$ and $S_2=\{y_1,…y_n\}$ one obtains a sample $S_3=\{w_1,…w_m\}$ where $w_i=+1$ if $x_i›y_i$ and $w_i=-1$ if $y_i › x_i$. The equal observations, called ties, are skipped. The test statistic T=# of + signs obeys the binomial distribution B(m, 0.5).For large values of m one may use the standard normal statistic Z :

$$Z = \frac{2T - m}{\sqrt{m}} \qquad (1)$$

Second test from this family is the McNemar test allowing to detect differences in pairs of qualitative variables (X,Y). The possible outcomes for (X,Y) are (0,0), (1,1), (1,0) and (0,1).The McNemar test consists in coding : (0,1) as +1 and (1,0) as −1, skipping other couples of observations as ties and performing the sign test described above.

Third test is the Cox and Stuart test. This test is a test for a trend in the time series of data. Given a sequence of data points: $x_1,x_2 …x_{2k}$ we pair the observations $(x_1, x_{k+1})$, $(x_2, x_{k+2})$ etc. If the first member of a couple is smaller than the second a couple gets a code *+1*. If the second member of the couple is smaller than the first one a couple gets a code *-1*. Next we perform the sign test described above.

### 4.2 The Runs tests

A run is a sequence of like elements that are preceded and followed by different elements or no elements at all. For example in a sequence 10110001001101110 we have 10 runs and in a sequence 11111100000 we have 2 runs. If the appearance of 1 or 0 is purely random the number of runs cannot be too small or too big. If the total sample size is $n=n_1+n_2$ the test statistics to test the randomness of the series is R=# of runs. The expectation value of R is:

$$E(R) = \frac{2n_1 n_2}{n_1 + n_2} + 1 \qquad (2)$$

Its variance is:

$$Var(R) = \frac{2n_1 n_2 (2n_1 n_2 - n_1 - n_2)}{(n_1 + n_2)^2 (n_1 + n_2 - 1)} \tag{3}$$

For large samples one may use standard normal test statistic:

$$Z = \frac{R - E(R)}{\sqrt{Var(R)}} \tag{4}$$

For small samples the p-values for the test may be found from the tables or using some statistical software.

The second test from the family is the Wald-Wolfowitz test in which the null hypothesis that two populations have the same distribution is tested. We arrange the values of two samples in increasing order in one sequence regardless of the population from which each is taken. Next we assign to the observations from the population II and I the codes 1 and 0 respectively. We obtain a sequence of zeros and ones on which we perform the run test described above.

*4.3.* The rank tests

In these tests we rank the observations from smallest to largest and then we use the ranks for the actual observations.

The most popular test from this family is Mann-Whitney U Test in which we test the null hypothesis that the distributions of two populations are identical. We combine two random samples $S_1$ and $S_2$ from these populations and rank all the observations. To all tied observations the same averaged rank is assigned. The Mann-Whitney U statistic is:

$$U = n_1 n_2 + \frac{n_1 (n_1 + 1)}{2} + R_1 \tag{5}$$

where $R_1$ is the sum of ranks scored by the observations from sample the $S_1$. The expectation value of U is:

$$E(U) = \frac{n_1 n_2}{2} \tag{6}$$

and the variance of U is:

$$Var(U) = \frac{n_1 n_2 (n_1 + n_2 + 1)}{12} \tag{7}$$

For large samples one can use the standard normal statistic:

$$Z = \frac{U - E(U)}{\sqrt{Var(U)}} \tag{8}$$

and the tables exist for the smaller samples.

The second test is the Wilcoxon signed-rank test testing the equality of the medians of two populations The observations from two samples of the same size n are paired and the differences between the first and the second member of each pair are recorded: $d_1, d_2, \ldots d_k$. Next the absolute values of these differences are ranked and the sums of the ranks $R_1$ and $R_2$ scored by the positive and the negative differences are found. The test statistic is: T= min $(R_1, R_2)$. The expectation value of T is:

$$E(T) = \frac{n\ (n\ +1)}{4} \qquad (9)$$

and the variance of T is:

$$Var(T) = \frac{n\ (n+1)(2n+1)}{24} \qquad (10)$$

For large samples one can use the standard normal statistic Z and the tables exist for the smaller samples.

The last test from this family is the Kruskal-Wallis Test. It is the generalization of the Mann-Whitney test to k populations. We start with k different samples $S_j$ of the size $n_j$, drawn from the k populations we want to compare. We combine all these samples in one sample of the size n=$n_1$+…+$n_k$ and we rank all the observations from the smallest to the largest. If $R_j$ is the sum of the ranks scored by the observations from the sample $S_j$ then the Kruskal-Wallis statistic is:

$$H = \frac{12}{n(n+1)}\left(\sum_{j=1}^{k}\frac{R_j{}^2}{n_j}\right) - 3(n+1) \qquad (11)$$

For large samples this statistic obeys approximately the chi-square distribution with k-1 degrees of freedom.

This finishes our review of non-parametric compatibility tests that can be easily used as purity tests.

### 4. Time series analysis

A time series is a set of measurements, which are ordered through time [43-46]. The simplest introduction is given in [45]. With a measurement at time t we associate a random variable $Z_t$ obtaining a family of random variables $\{Z_t\}$. To detect a fine structure in the time series only usually uses various models of $Z_t$ trying to match the existing data. In this section we assume that the time is discrete.

*5.1.* The additive model:

$$Z_t = T_t + S_t + C_t + I_t \qquad (12)$$

where T is the trend component , S is the seasonal component, C is the cyclical component and I is the irregular component of the series. The seasonality is studied in terms of the multivariate regression with dummy variables.

*5.2.* The multiplicative model:

$$Z_t = (T_t)(S_t)(C_t)(I_t) \qquad (13)$$

The seasonality is studied in this model by so called ratio-to moving average method. A moving average of a time series in the average of a fixed number of observations that moves as we progress down the series. A moving average eliminates seasonal variability of the data. The series of the moving averages gives the information about the trend and the cyclic components. If we divide each observation in a time series by the moving average of its subgroup of observations we obtain *a ratio to moving average series* which, if the multiplicative model is correct, helps us to study the seasonality and the irregular component.

If the various components are present in the series we can fit some parametric models to the experimental data, make the forecasts and check them with the new data.

*5.3 Exponential smoothing methods*

Exponential smoothing is a forecasting method where the forecast is based on a weighted average of current and past observations. The weights decline geometrically as we go back in time. One parameter exponential smoothing model can be given by the following recursive formula:

$$\hat{Z}_{t+1} = wZ_t + (1-w)\hat{Z}_t \tag{14}$$

where $\hat{Z}_t$ is a forecast value of the variable Z for time t , $Z_t$ is the actual measured value of the variable Z at time t and w is a smaller than 1 positive fraction called the weighting factor.

*5.4. The Box Jenkins methodology.*

In Box Jenkins methodology [48,49] first we hypothesize an appropriate statistical model, then we estimate the model parameters and test the model adequacy and finally we use it for forecasting. Let us list these steps in more detail:

1. Identify one or more models that describe the time series well. The identification is done statistically by testing hypotheses about the correlation structure of the series
2. Estimate parameters using the linear, non linear or lagged –variable regression.
3. Conduct model diagnostic and select the best among the models identified in the step 1.
4. Use the model and if it is not working well go back to the steps 1 and/or 2 and find a better model.

The theoretical models used in the step 1 are so called autoregressive integrated moving average (ARIMA) processes discussed in detail in [49].

6.Conclusions.

The perfect randomness and predictable completeness of QT have been taken for granted by the fathers of QT and by the majority of the physical community. The intrinsic randomness and completeness were understood in the following sense: *If some discrete physical observable is measured on the physical systems prepared in the same way and k different outcomes are possible then the statistical distribution of these outcomes is completely described by a set of the corresponding probabilities provided by QT.*

The problem of the completeness was considered for a long time to be metaphysical since it was believed that even if there existed a successful hidden variable theory it should reproduce all statistical predictions of QT. The problem of completeness changed from metaphysical to experimental when Bell [16] showed that a large class of so called local and realistic hidden variable (LRHVM) models constructed to explain the long range correlations in spin polarization correlation experiments (SPCE) led to the so called Bell or CHSH inequalities which were violated by the predictions of QT for some direction of polarization analysers [17,20]. The violation of Bell inequalities in SPCE confirmed the contextual character of QT and eliminated a possibility of the description of SPCE in terms of LRHVM but by no means provided the proof of the completeness of QT {35,4-8]

As is it was explained in the previous sections only the purity tests and a careful analysis of the time series of data may prove or disprove the irreducible randomness of the individual measurements and the predictable completeness of QT. The purity tests are simple and can be performed by any unit of the experimental group responsible for the statistical analysis of the data. The systematic analysis of the time series described in the section 5 is much more difficult [46-49] and to be conclusive should be done with the help of the statisticians specialized in time series and forecasting.

The purity tests may give additional arguments in favour of the statistical interpretation of the theory showing that some quantum ensembles believed to be pure are in fact imperfect mixtures with respect to some uncontrolled parameters describing the invisible sub-phenomena. It would give therefore the arguments against the instantaneous reduction of the wave packets and made impossible the treatment of the quantum state vectors as the attributes of the individual particles [4,36]

The results of the purity test may give the indication that QT is not predictably complete. To make a decisive proof one has to analyze experimental time series using the methods of references [46-49] in order to find some reproducible fine structures in the data, which were not predicted by QT. Such a discovery would be a revolution. We hope that perhaps some physicists will get encouraged to do these tests which do not require millions of dollars, and which can improve the understanding of quantum mysteries and to find the limitations of QT.